\documentclass[MSNbibl,dvips]{arxstspdf}
\usepackage{flushend}
\usepackage{stfloats}
\usepackage{graphicx}

\volume{26}
\issue{4}
\pubyear{2011}
\firstpage{608}
\lastpage{612}
\doi{10.1214/11-STS356REJ}
\referstodoi{10.1214/11-STS356}

\begin{document}
\begin{frontmatter}
\vspace*{6pt}
\title{Rejoinder}
\runtitle{Rejoinder}

\begin{aug}
\author[a]{\fnms{Jelle J.} \snm{Goeman}\corref{}\ead[label=e1]{j.j.goeman@lumc.nl}}
\and
\author[b]{\fnms{Aldo} \snm{Solari}\ead[label=e2]{aldo.solari@unimib.it}}
\runauthor{J. J. Goeman and A. Solari}

\affiliation{Leiden University Medical Center and University of Milano-Bicocca}

\address[a]{Jelle J. Goeman is Associate Professor, Department of Medical Statistics
and Bioinformatics (S5-P),
Leiden University Medical Center, P.O. Box 9600, 2300 RC Leiden, The
Netherlands \printead{e1}.}
\address[b]{Aldo Solari is Assistant Professor, Department of Statistics, University of
Milano-Bicocca, Italy \printead{e2}.}

\end{aug}



\end{frontmatter}

We are thankful to the three discussants for their helpful and
stimulating comments to our work. We value especially the many
suggestions for extensions of the methodology given by all three discussants.

One such suggested extension, with which we were very pleasantly
surprised, was Meinshausen's discussion of the defining hypotheses. The
defining hypotheses of a closed testing result, as we presented them,
describe the result of the closed testing procedure as a union of
intersections of hypotheses. Meinshausen's suggestion is to rewrite the
same result as an intersection of unions, which can always be done with
some basic algebra. We like to call the resulting collection the \emph
{shortlist}, because it shortlists the candidate combinations of false
hypotheses: with $1-\alpha$ confidence at least one of the shortlist
sets is a subset of the actual set of false hypotheses. Rewriting the
result in this way gives a surprisingly complementary perspective on
the results of the procedure, that is intuitive and can be very helpful
for interpretation of the test procedure's results, as demonstrated by
Meinshausen. We have added the possibility to calculate the shortlist
to the \emph{cherry} package, with thanks.

An interesting variant to our procedure was suggested by Heller. In
this variant not just the elementary hypotheses are candidates for
validation, but also (sub)families of hypotheses are of interest by
themselves. Such subfamilies of hypotheses can be represented by
intersection hypotheses in the closure. An example of this can be a
genomic data analysis in which both single genes and gene sets are of
interest.\vadjust{\goodbreak} It is \mbox{relatively} easy to extend the methodology to allow
confidence statements on $\#(\mathcal{R} \cap\mathcal{T})$ where
$\mathcal{R}$ is a collection of index sets representing intersection
hypotheses that are chosen as candidates for validation. Confidence
bounds $t_\alpha(\mathcal{R})$ can be derived using similar reasoning
as used to find the familiar $t_\alpha(R)$.

The link between our proposed approach and the partial conjunction
approach of Heller is a strong one, which we acknowledge and to which
we should perhaps have pointed more explicitly. In our notation, the
partial conjunction hypothesis $H^{u/n}$ is given by
\[
H^{u/n} = \bigcup_{I \in\mathcal{C}^u} H_I,
\]
where $\mathcal{C}^u = \{C \in\mathcal{C}\dvtx |C|=u\}$. This is exactly
the union of hypotheses that has to be rejected in the closed testing
procedure to be able to conclude that $t_\alpha(R) < u$ for $R=\{
1,\ldots,n\}$. Our method can be seen as extending upon Heller's work
by allowing other choices of $R$, but diverging from it where the issue
of multiple testing of partial conjunction hypotheses is concerned, a
problem which Benjamini and Heller (\citeyear{Benjamini2008}) addressed in an FDR context. In her
discussion, Heller introduces the partial conjunction hypothesis
$H^{u/|R|}$, which we would prefer to denote $H^{u/R}$ because it
depends on the set $R$, not just on its cardinality, and formally
define it
\[
H^{u/R} = \bigcup_{I \in\mathcal{C}^u_R} H_I,
\]
where $\mathcal{C}^u_R = \{I \subseteq R\dvtx  |I|=u\}$. This is indeed a
central hypothesis to the approach we presented, which can
alternatively be described as simultaneously testing~$H^{u/R}$ for all
sets $R \subseteq\{1,\ldots,n\}$ and for all $1\leq u\leq|R|$.
Thinking of the procedure in terms of such partial conjunctions is a
valuable perspective on our procedure. We also value Heller's concept
of sufficient combining functions, which promises to be useful for
finding new shortcuts and proving their validity.

The computational problems associated with\break closed testing have been
rightly stressed by both Meinshausen and Westfall, and we are aware of
these problems. We cannot stress enough that shortcuts are crucial\vadjust{\goodbreak} for
the usability of the method we have proposed unless the number of
tested hypotheses is small. However, we think that the shortcuts we
have described in our paper are only a beginning, and that many more
shortcuts are possible. Of practical relevance to genomics research are
especially shortcuts of the types discussed in Section 4.4 of our
paper, in which a limited number of intersection hypotheses are tested
with a non-consonant test, while the rest of the hypotheses can be
tested using weighted Bonferroni-based combinations of these test
results. Such shortcuts are relatively easy to design and they can be
tailored to the specific needs of practical testing problems. Other,
more general shortcuts are likely to be found as well.

Some of the issues raised by the discussants require a somewhat more
thorough discussion or give rise to some interesting elaborations of
the theory. We will take the opportunity to go into a few subjects more
deeply, elaborating on the issue of power, mentioned by Meinshausen and
Westfall, on the issues of restricted combinations and adjusted
$p$-values, both discussed by Westfall, and finally on the complicated
practice of exploratory research, as commented on by Heller.

\section{Power of the Proposed Approach}

Both Meinshausen and Westfall commented on the power of our proposed
procedure, illustrating their points with example data and simulations.
Pow\-er is a crucial consideration not just in confirmatory, but also in
exploratory settings. It is difficult, however, to talk about the power
of our method in a~general way, because the closed testing procedure
that underlies it is extremely versatile. The power properties of the
procedure depend crucially on the power properties of the chosen local
test. We will illustrate this by looking at the examples given by the
two discussants in more detail.

Westfall analyzes the famous Golub et~al. (\citeyear{Golub1999}) microarray dataset using
both Fisher combinations and Bonferroni as a local test, the latter
leading to Holm's (\citeyear{Holm1979}) procedure. On a familywise
error of 0.05, Holm finds 37 out of 7,129 elementary hypotheses to be
individually significant, whereas Fisher combinations do not find any
significant elementary hypotheses. Taking into account that Fisher
combination tests are known to be anti-conservative in these data due
to correlations among the test statistics, this comparison does indeed
seem to come out clearly in favor of Bonferroni/Holm. This assessment\vadjust{\goodbreak}
changes, however, if we try to make a statement that is not of
familywise error type, such as counting how many false hypotheses are
present\break among the 7,129 hypotheses. The procedure based on Bonferroni
states at 95\% confidence that the 37 hypotheses found with familywise
error control are false, but can say no more than that. The method
based on Fisher combinations, on the other hand, although it could not
confidently point to any individual hypothesis as false, finds with
95\% confidence that no fewer than 1,828 false hypotheses are present
among the 4,082 hypotheses with smallest $p$-values. This example
illustrates that different local tests result in procedures with
completely different properties. Procedures based on highly consonant
tests, such as Bonferroni's, tend to have good power for intersection
hypotheses $H_I$ of low cardinality $|I|$ and, consequently, are the
method of choice for familywise error statements. Procedures based on
highly non-consonant tests, such as Fisher combinations, tend to have
good power for intersection hypotheses~$H_I$ of high cardinality $|I|$,
and, consequently, typically give superior bounds $t_\alpha(R)$ for
large sets~$R$. It is interesting to note that the Simes local test
takes an intermediate course for this dataset, finding the same 37
hypotheses to be significant from a~familywise error perspective, but
finding 111 additional false null hypotheses in total among the 7,129
genes due to its additional non-consonant rejections.

\begin{figure}[t]

\includegraphics{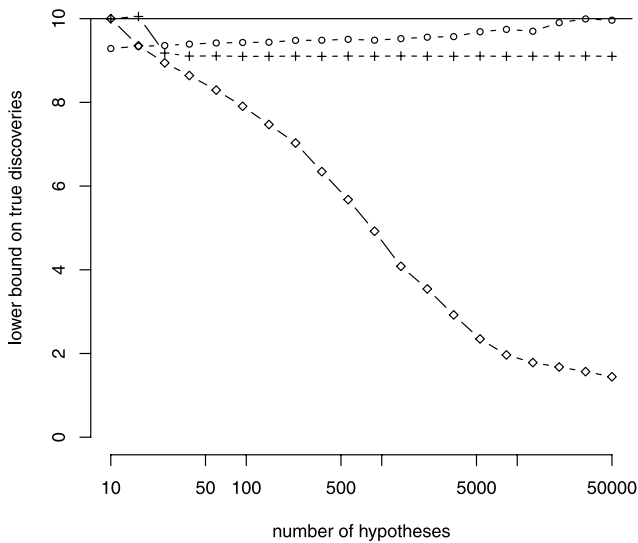}

\caption{Simulation study as in Figure 1(\textup{b}) in Meinshausen's
discussion, but with Simes' local tests added. Meinshausen's method
(circles), Fisher combinations (diamonds) and Simes (plusses).} \label{simulation}
\end{figure}

Related remarks can be made about the simulation study performed by
Meinshausen. His simulated alternatives with a large number of
hypotheses $m$ have a very sparse but strong signal. This is a kind of
signal that Fisher combinations are not very good at detecting, as is
clearly illustrated by the simulation results. The observed low power
in the simulation is more a feature of Fisher combinations as a~local
test than of the confidence set method as such. If a sparse but strong
signal was expected, Fisher combinations should not have been chosen as
the local test. If we redo the simulation with Simes local tests we get
a completely different picture (Figure~\ref{simulation}), with comparable
power to Fisher combinations for low values of $m$, and only slightly
lower power compared to Meinshausen (\citeyear{Meinshausen2006}) for large values of~$m$.

Depending on the alternative that is to be picked up, and depending on
the type of statements that are to be made, different choices of local
tests may lead to procedures with good or bad power. Obviously, this
gives much room for the development of powerful procedures tailored to
specific research questions on specific types of data.

\section{Restricted Combinations}

Westfall raises the important issue of restricted combinations.
Restricted combinations occur when, because of logical relationships
between hypotheses, the collection $T$ of true null hypotheses is a
priori restricted, and some elements of the closure cannot be equal to
the true set. In the example given by Westfall, the hypotheses are
$H_1\dvtx \mu_1=\mu_2$, $H_2\dvtx \mu_1=\mu_3$ and $H_3\dvtx \mu_2=\mu_3$. For
these hypotheses, $T=\{1,2\}$ is not possible, as simultaneous truth of
$H_1$ and $H_2$ implies truth of $H_3$. Similarly, any other set $T$ of
cardinality 2 is excluded, and $T$ can only take the values
$\varnothing
$, $\{1\}$, $\{2\}$, $\{3\}$ or $\{1,2,3\}$. We call those sets that
cannot be equal to $T$ \emph{incongruent}. There is an immense body of
literature on multiple testing in the presence of restricted
combinations, starting with the famous paper of Shaffer (\citeyear{Shaffer1986}), but
we did not consider this issue in our paper.

Westfall claims that the method we have proposed may be conservative if
restricted combinations are present. This is true. However, a very
simple extension of the method can remove this conservativeness in a
general way, which is very similar to Westfall's treatment of the
specific example. This extension follows from the Sequential Rejection
Principle (Goeman and Solari, \citeyear{Goeman2010}). Applied to closed testing, this principle
states that the local test of each $H_I$, $I \in\mathcal{C}$, when it
is its turn to be tested, may assume that all hypotheses $H_J$, $J
\supset I$ are false. For an incongruent\vadjust{\goodbreak} set $I$, falsehood of all such
hypotheses $H_J$ immediately implies that $H_I$ itself is false.
Therefore, even the test that always rejects is a valid local test for
the incongruent set $I$. Consequently, in the presence of restricted
combinations, we may assume that the local test always rejects all
incongruent sets.

The same conclusion may also be arrived at in an alternative way, by
using the partitioning principle (Finner and Strassburger, \citeyear{Finner2002}) rather than closed
testing to make the confidence sets. For each $I \in\mathcal{C}$, let
the corresponding partitioning hypothesis be
\begin{equation} \label{partitioning}
J_I = \bigcap_{i \in I} H_i \setminus\bigcup_{j \notin I} H_j.
\end{equation}
Suppose an $\alpha$-level test is available for every $J_I$, $I \in
\mathcal{C}$, and let $\mathcal{V}\subseteq\mathcal{C}$ be the index
set of the partitioning hypotheses rejected by their corresponding
test. Then, analogously to the closed testing based procedure, by the
partitioning principle the intersection hypothesis $H_I$, $I \in
\mathcal{C}$, is rejected whenever $J \in\mathcal{V}$ for every $J
\supseteq I$. From this, we can make the set $\mathcal{X}$ of rejected
intersection hypotheses and derive the upper confidence limit $t_\alpha
(R)$ as before. The whole procedure is completely analogous to the one
presented in the paper, only the smaller partitioning hypotheses~$J_I$,
$I\in\mathcal{C}$, take the role of the closure hypotheses~$H_I$ when
finding the set $\mathcal{X}$. As $J_I \subseteq H_I$, for every~$I$,
any valid test of $H_I$ is also a valid test for $J_I$, but $J_I =
\varnothing$ for every incongruent $I$, so that we can safely reject
every incongruent $J_I$. Consequently, again, we may assume that
incongruent hypotheses are always rejected by their local test.

If we extend our method in this way, we have a~general solution for the
problem of restricted combinations. Using this extension, the
conservativeness noted in Westfall's example disappears, and the
stronger statements he obtained are recovered. For concrete examples of
families with restricted combinations, finding good shortcuts that take
restricted combinations into account may, of course, still be a~challenging problem.

\begin{figure*}[t]

\includegraphics{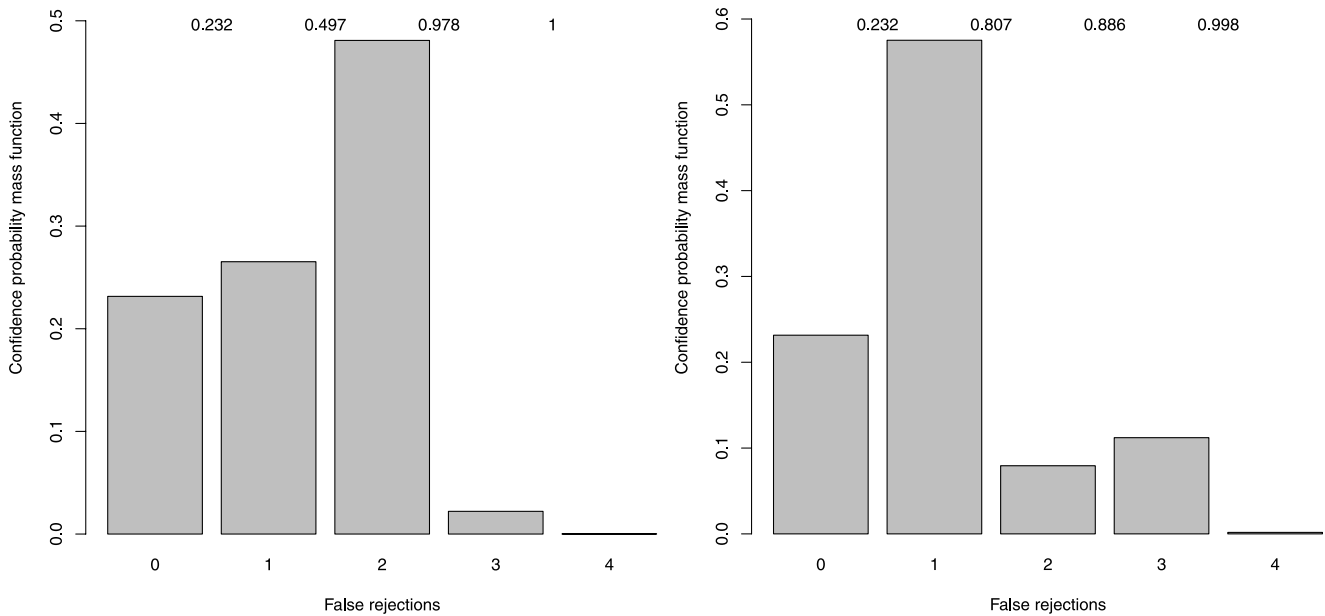}

\caption{Confidence probability mass function for $R= \{$waist,
forearm, calf, thigh\} (left-hand side) and $R= \{$waist, forearm,
height, thigh\} (right-hand side). Cumulative percentages are given at
the top of the figure.} \label{fiducial1}\vspace*{-3pt}
\end{figure*}

\section{Adjusted $\lowercase{p}$-Values}

The adjusted $p$-value is an important feature of multiple testing
procedures, that conveys valuable additional information over the
simple decision to reject or not to reject, as Westfall rightly points
out. We did not go into the issue of adjusted $p$-values in the paper
because we wanted to stress the analogy with confidence intervals,
which are typically calculated for a fixed $\alpha$. It is possible,
however, to find an analogy to adjusted $p$-value that conveys the same
type of additional information, or even more.

By definition, an adjusted $p$-value of a certain hypothesis is the
smallest $\alpha$-level that allows rejection of that hypothesis. By
giving the threshold level $\alpha$ that distinguishes those $\alpha
$-levels for which the hypothesis would or would not be rejected, the
adjusted $p$-value gives the information what inferences would have
been obtained if some other value of~$\alpha$ would have been chosen.
Analogously, in the exploratory setting, we may also vary the value of
$\alpha$ and plot the upper confidence bound $t_\alpha(R)$ of the
number of true hypotheses $\tau(R)$ as a function of $\alpha$. Just
like the adjusted $p$-value, this shows the dependence of our
conclusions on the arbitrary choice of~$\alpha$.\looseness=-1

An intuitive way to visualize the dependence\break of~$t_\alpha(R)$ on $\alpha
$ is through a plot analogous to a \emph{confidence distribution}
(Singh, Xie and Strawderman, \citeyear{Singh2007}). This can be obtained by interpreting the plot of $t_\alpha
(R)$ as a function of $\alpha$ as if it was the quantile function of
some discrete probability distribution, and plotting the differential
of that, that is, the associated ``probability mass function.'' For the
sets $R=$ \{waist, forearm, calf, thigh\} and $R=$ \{waist, forearm,
height, thigh\} in the example of the physical data of Section~3 in the
paper, this plot is given in Figure~\ref{fiducial1}. From this plot, we
can read off the 95\%\ confidence limit $t_{0.95}(R)$ and the estimate
$t_{1/2}(R)$ by finding the 95th quantile and the median, respectively.
We can also read off the familywise\vadjust{\goodbreak} error adjusted $p$-value at 1 minus
the confidence distribution at 0.

It is important to realize that a confidence distribution is a random
variable, not a probability distribution, in the same way that an
adjusted $p$-value is a random variable, not a probability. The
representation as if it was a distribution should just be seen as a
convenient way to visualize the dependence of~$t_\alpha(R)$ on $\alpha
$. It is the direct analogue of the adjusted $p$-value for the method
we have proposed.

\section{The Practice of Exploratory Research}

The methods we have presented still require a~quite formal and planned
design in which hypotheses have been formulated before data analysis
and tests for intersection hypotheses are chosen beforehand. Such a way
of working is close enough to actual data analysis in many genomics
experiments, but it is far too formal to capture the great variety and
freedom of true exploratory research. In actual practice, researchers
often first perform a goodness-of-fit test on the same data before
deciding what test to do. Researchers typically do additional unplanned
hypotheses tests to detect the presence of effects suggested by plots
of the same data. Also, researchers sometimes perform additional tests
as a consequence of the nonsignificance of other test, because they
are not satisfied\vadjust{\goodbreak} with the non-significant result obtained. It is
impossible to capture the true complexity of exploratory research in
any formal method.

Heller suggests a two-stage approach that separates the data used for
exploratory analysis in two parts. The first part is used as a pilot
merely to decide what an appropriate model would be, and how
intersection hypotheses should be tested, in a~completely free
exploratory manner. The end result of this data analysis would be a
list of hypotheses and a plan for the closed testing procedure to be
used in the next exploratory phase, which is then formal enough to
allow use of the methods we have proposed. This proposal is practical
and elegantly simple, mimicking the data splitting between exploratory
and confirmatory research. It is also good that it stresses the need
for a pilot experiment, which can be useful in many other respects as
well. A practical problem, however, may be that many crucial decisions,
regarding, for example, which test can be expected to have most power
or which model fits best, may require quite large sample size, so that
relatively small pilot experiments may not be adequate. One can also
object philosophically to the idea, saying that the same arguments that
can be used to change the empirical cycle from a two-phase process into
a three-phase one could again be used to add a new fourth initial phase
to the cycle, because the methods used in the pilot phase may be wrong
or lacking in power. This way there would be no end to data
splitting.

In the end, we think exploratory research in its most general form is
too fluid to be captured in formal methodology. We feel, however, that
we have demonstrated that more things are possible in the exploratory
context than was commonly thought, and we hope we have stimulated the
discussion on multiple testing in exploratory research. In our turn, we
have been greatly stimulated by the contributions of the three
discussants, for which we are thankful.


\end{document}